\pdfoutput=1
\documentclass{article}
\usepackage{spconf,amsmath,graphicx}
\usepackage{array}
\usepackage{booktabs}
\usepackage{amssymb}
\usepackage{graphicx}
\usepackage{setspace}
\usepackage{url}
\providecommand{\tabularnewline}{\\}

\title{Desensitized RDCA Subspaces for Compressive Privacy in Machine Learning}
\name{Artur Filipowicz \qquad Thee Chanyaswad \qquad S.Y. Kung\thanks{Thanks to the Brandeis Program of the Defense Advanced Research Project Agency (DARPA) and Space and Naval Warfare System Center Pacific (SSC Pacific) under Contract No. 66001-15-C-4068 for funding support, and to Professor J. Morris Chang from Iowa State University for invaluable discussion
and assistances.}}
\address{Princeton University\\Princeton, NJ}

\begin{document}
\ninept
\maketitle
\begin{abstract}
The quest for better data analysis and artificial intelligence has
lead to more and more data being collected and stored. As a consequence,
more data are exposed to malicious entities. This paper examines the
problem of privacy in machine learning for classification. We utilize
the Ridge Discriminant Component Analysis (RDCA) to desensitize data
with respect to a privacy label. Based on five experiments, we show
that desensitization by RDCA can effectively protect privacy (i.e.
low accuracy on the privacy label) with small loss in utility. On
HAR and CMU Faces datasets, the use of desensitized data results in
random guess level accuracies for privacy at a cost of 5.14\% and
0.04\%, on average, drop in the utility accuracies. For Semeion Handwritten
Digit dataset, accuracies of the privacy-sensitive digits are almost
zero, while the accuracies for the utility-relevant digits drop by
7.53\% on average. This presents a promising solution to the problem
of privacy in machine learning for classification.
\end{abstract}
\begin{keywords}
Compressive Privacy, Ridge Discriminant Component Analysis (RDCA), Privacy-preserving Data Mining/Machine Learning, Data Desensitization, Dimension Reduction
\end{keywords}

\section{Introduction}

Innovation in the 21\textsuperscript{st} Century electronics centers
around data processing. Progress is fueled by the symbiotic relationship
between big data and machine learning in which machine learning allows
us to interpret big data and big data allows us to train large machine
learning models. In the world of big data, videos, photos, emails,
banking transactions, browsing history, GPS tracks, and other personal
data are continuously collected and stored by organizations for analysis.
These data may be circulated around the Internet without the data
owner's knowledge and be at risk of exposure to malicious entities.
A few recent data leakages are described by \cite{dataBreach}, and
many other possible attacks on privacy have been reported or proposed
\cite{netflix_prize,RefWorks:181,RefWorks:182,RefWorks:183}.

The complete problem of maintaining privacy is complex. It is distributed
temporally since data owner's present and past actions can compromise
privacy. It is distributed spatially as the data owner has personal
information in multiple accounts, devices, and physical locations.
Our focus is privacy protection in the context of machine learning
for classification, at the time and location the data owner, the user,
submits his/her data to a machine learning service, the server. 

A classical solution to this problem is encryption. The user encrypts
his/her information before submission, and the server decodes the
submitted data. However, the server may leak these data to malicious
entities. Therefore, it should not be trusted and should not receive
information which compromises privacy. In machine learning for classification,
this is information which maximizes the classification accuracy of
the utility label while minimizing the classification accuracy of
the privacy label. Several different ideas have been proposed to attack
this problem such as noisy data reconstruction \cite{RefWorks:89,RefWorks:90},
rotational and random perturbation \cite{RefWorks:93,RefWorks:94},
microaggregation of data \cite{RefWorks:155}, privacy-centric classifier
designs \cite{RefWorks:98,RefWorks:103,RefWorks:154,chanyaswad2017compressive},
etc.

Our method is based upon Compressive Privacy \cite{RefWorks:184,ACM2016,mlsp2016}
approach to this problem. It utilizes the concept of data desensitization
\textendash{} modifying the data by reducing the number of features
such that the privacy is protected. We employ Ridge Discriminant Component
Analysis (RDCA) \cite{kung2015discriminant,RefWorks:33,ACM2016} to
desensitize data before they are submitted to a server. By deriving
the RDCA components with respect to the privacy label, two subspaces
are attained \textendash{} the privacy signal and privacy noise subspaces.
The privacy noise subspace is the subspace which has minimal classification
power with respect to the privacy label. Therefore, the proposed method
utilizes this privacy noise subspace to project the data onto in order
to desensitize the data. Even if the sever leaks the desensitized
data, a malicious entity cannot use these data to classify the user
under the privacy label. 

Based on the properties of the utility and privacy labels, we define
three problem classes \textendash{} Common-Unique Privacy, Common-Common
Privacy and Split-Label Privacy problems. To test the potency of our
method on our three different classes, we present an example dataset
\textendash{} HAR, CMU Faces, and Semeion Handwritten Digit \textendash{}
for each class and show that desensitization by RDCA can effectively
protect privacy by reducing the privacy accuracy to the random guess
level in the HAR and CMU Faces datasets, and to almost zero on the
privacy-sensitive digits in the Semeion Handwritten Digit dataset.
On the other hand, the utility accuracies only drop by 5.14\% in the
HAR dataset, on average by 0.04\% in the CMU Faces dataset, and on
average by 7.53\% on the utility-relevant digits in the Semeion Handwritten
Digit dataset. This confirms that the proposed desensitization method
by RDCA is promising in providing a solution in privacy-preserving
machine learning.

\section{Privacy Problem Classes}

In a standard classification problem, for a given set of supervised
training data of $N$ samples and $M$ features, $\{X,\vec{y}\}$,
a classifier is trained to predict $\vec{y}$ based on $X$. When
considering privacy, we define a privacy label $\vec{y}^{(p)}$ and
a utility label $\vec{y}^{(u)}$. Our objective is then to minimize
the possibility of $\vec{y}^{(p)}$ being predicted based on $X$
while maximizing our classifier's ability to predict $\vec{y}^{(u)}$.
Based on this approach, we define three classes of problems. 

To define these problem classes, we first introduce the ideas of a
common and unique label. A unique label has a different class for
each user, for example social security number. A common label has
classes which are shared by multiple users, eye color being an example. 

The three problems we define are Common-Unique Privacy, Common-Common
Privacy, and Split-Label Privacy. 
\begin{itemize}
\item In Common-Unique Privacy problem, $\vec{y}^{(u)}$ or $\vec{y}^{(p)}$
is a common label while the other label is unique. An application
example we explore is human activity recognition. In this example,
the unique label is the identity of the user submitting activity data,
while the common label is the type of activity (walking, running,
etc.) performed by the user. 
\item In Common-Common Privacy problem, both $\vec{y}^{(u)}$ and $\vec{y}^{(p)}$
are common labels. An example we examine is facial feature recognition.
Specifically, for each image, there are two common labels indicating
if the user is wearing sunglasses and his pose. 
\item Lastly, in Split-Label Privacy problem, $\vec{y}^{(u)}$ and $\vec{y}^{(p)}$
are derived from a single label $\vec{y}$ where some classes are
grouped. Our example is Optical Character Recognition where we wish
to recognize digits 0 to 4 while protecting digits 5 to 9 from being
recognized. The digits may be responses on a survey about marriage
where 0 to 4 represents single-never-married, single-divorced, single-widowed
etc. and digits 5 to 9 represent married-male-female, married-male-male,
etc. In this case, the response between 5-9 may leak private information
about sexual orientation, and therefore should not be uniquely identifiable,
unlike the utility digits 0-4, of which unique identifiability may
be useful for marketing purposes.
\end{itemize}

\section{Desensitization by RDCA Subspace Projection}

\subsection{Ridge Discriminant Component Analysis (RDCA)}

\begin{figure}
\begin{centering}
\includegraphics[width=1\columnwidth]{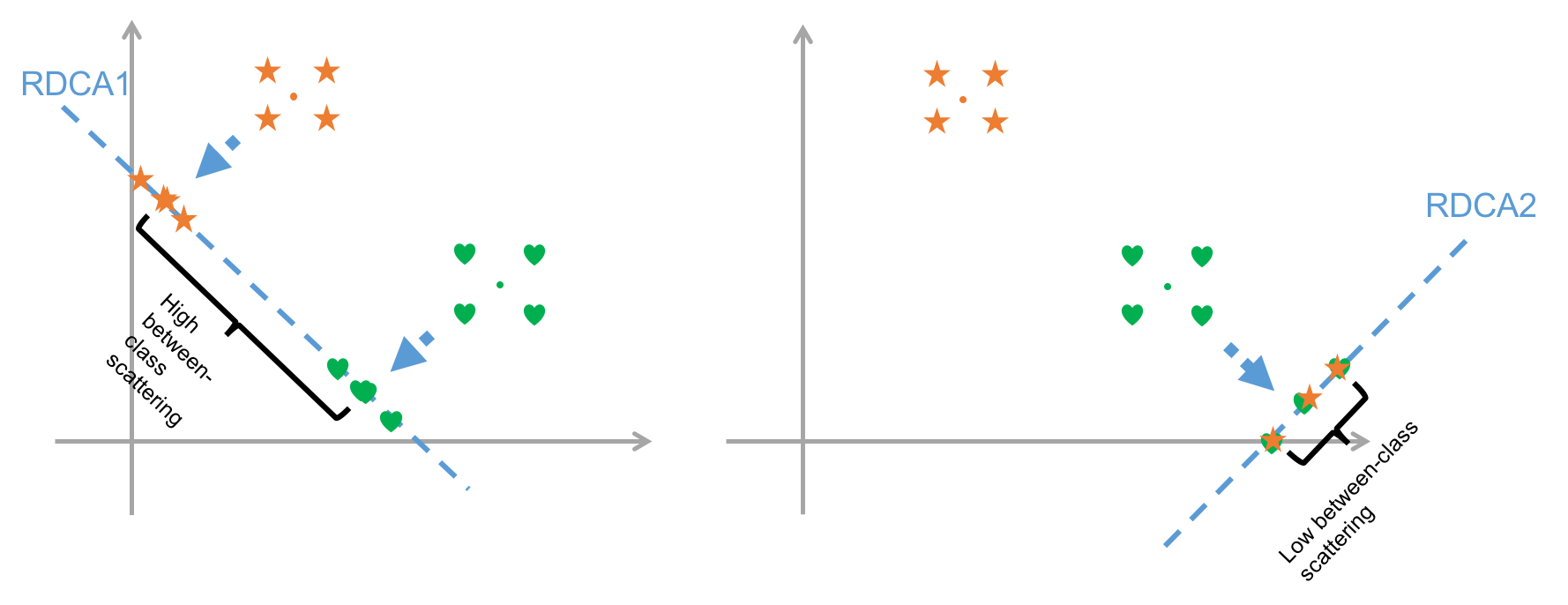}
\par\end{centering}
\caption{Illustration for the RDCA signal and noise subspaces with $M=2$ and
$L=2$. In the signal subspace (left), the distance between class
centroids is high, where as in the noise subspace (right), the distance
between class centroids is low relative to the distances among all
of the samples. \label{fig:Illustration_DCA}}
\end{figure}

Ridge Discriminant Component Analysis (RDCA) \cite{kung2015discriminant,RefWorks:33,ACM2016}
aims at finding the subspace that maximizes the discriminant distance
among the classes. Given an $L$-class classification problem, RDCA
is able to provide the $(L-1)$-dimensional subspace where all discriminant
power lies, and the remaining subspace where no discriminant power
remains. Conceptually, RDCA aims at maximizing the ratio between between-class
scattering (signal) and total scattering (total power). Hence, the
noise subspace from RDCA corresponds to the subspace where the distance
scattering among samples is comparable to the distance scattering
among centroids of the classes. This phenomenon is illustrated in
Figure \ref{fig:Illustration_DCA}. For the mathematical discussion
and derivation of RDCA, we refer the readers to \cite{kung2015discriminant,RefWorks:33,ACM2016}. 

The essential property that is relevant to this proposed work is the
fact that RDCA has the capability to provide the signal and noise
subspaces with respect to a label. Given the discriminant components
derived from RDCA, $\{\vec{w}_{1},\vec{w}_{2},\ldots,\vec{w}_{L-1},\vec{w}_{L},\ldots,\vec{w}_{M};\vec{w}_{i}\in\mathbb{R}^{M}\}$,
as ordered by the decreasing discriminant power, the signal and noise
subspaces are therefore defined as $span(\{\vec{w}_{1},\vec{w}_{2},\ldots,\vec{w}_{L-1}\})$,
and $span(\{\vec{w}_{L},\vec{w}_{L+1},\ldots,\vec{w}_{M}\})$, respectively. 

\subsection{Desensitized Subspace and Desensitized Data}

\begin{figure}
\begin{centering}
\includegraphics[width=1\columnwidth]{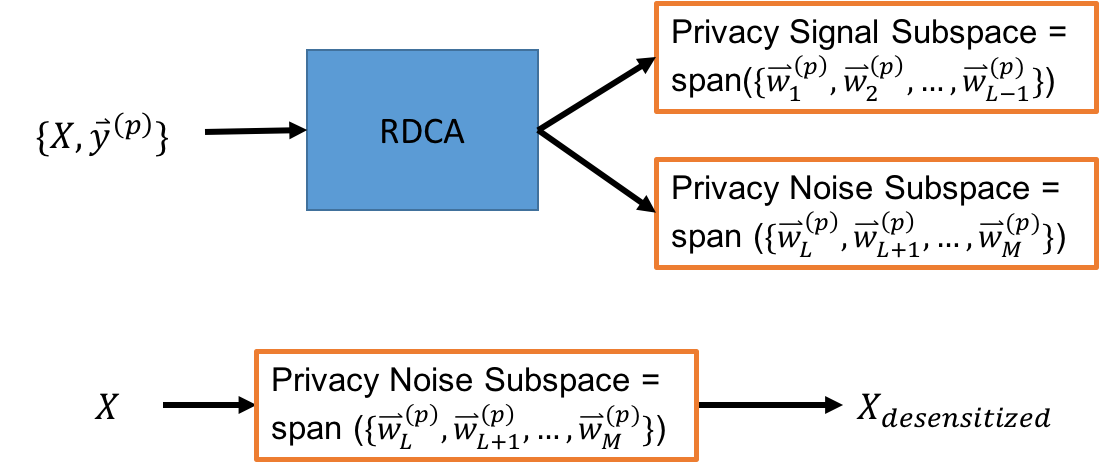}
\par\end{centering}
\caption{Illustration of the data subspace desensitization process. First,
the signal and noise subspaces spanned by the RDCA components are
derived from the data and the privacy label (top). Then, the data
are projected onto the privacy noise subspace to obtain the desensitized
data (bottom). \label{fig:privatization_process}}
\end{figure}

As RDCA can separate the signal and noise subspaces with respect to
a label, it lends itself nicely to the application of data desensitization.
By using the privacy label $\vec{y}^{(p)}$ to train RDCA, the privacy
signal subspace, $S_{S}^{(p)}=span(\{\vec{w}_{1}^{(p)},\vec{w}_{2}^{(p)},\ldots,\vec{w}_{L-1}^{(p)}\})$,
and the privacy noise subspace, $S_{N}^{(p)}=span(\{\vec{w}_{L}^{(p)},\vec{w}_{L+1}^{(p)},\ldots,\vec{w}_{M}^{(p)}\})$,
can be derived. Thus, by using only the privacy noise subspace for
the submitted data, the discriminant power of the privacy classes
can be minimized. Appropriately, the privacy noise subspace is called
the \emph{desensitized subspace,} and the data projected onto this
subspace are referred to as the \emph{desensitized data}. The procedure
for producing the desensitized data is summarized in Figure \ref{fig:privatization_process}.

\section{Experiments}

To provide real world examples of the three privacy problem classes
and show that RDCA can be used to protect privacy, we conducted experiments
on the Human Activity Recognition Using Smartphones, CMU Faces, and
Semeion Handwritten Digit datasets. For all experiments, SVM is used
as the classifier for both utility and privacy, and in the training
phase, cross-validation is used to tune the parameters.

\subsection{HAR}

Human Activity Recognition Using Smartphones (HAR) dataset \cite{RefWorks:169}
aims at using mobile sensor signals (accelerometer and gyroscope)
to predict activity being performed. The feature size of the dataset
is 561. The data are collected from 19 individuals performing six
activities. The dataset consists of 5379 samples for training and
798 samples left out for testing. The activity is defined to be the
utility, $\vec{y}^{(u)}$, whereas the person identification is defined
to be the privacy, $\vec{y}^{(p)}$. 

\subsection{CMU Faces }

CMU Faces dataset contains 640 grayscale images of 20 individuals
\cite{cmuFaces}. For each individual there is an image for every
combination of pose (straight, left, right, up), expression (neutral,
happy, sad, angry), and sunglasses (present or not). Images of size
32 by 30 pixels are used. Two experiments are performed:
\begin{itemize}
\item In the first experiment, the utility, $\vec{y}^{(u)}$, is defined
to be the pose and the privacy, $\vec{y}^{(p)}$, is the sunglasses
indicator. 
\item In the second experiment, the utility, $\vec{y}^{(u)}$, is defined
to be the sunglasses indicator and the privacy, $\vec{y}^{(p)}$,
is the pose.
\end{itemize}

\subsection{Semeion Handwritten Digit }

The Semeion Handwritten Digit dataset contains 1593 handwritten digits
from around 80 individuals. Every individual wrote each digit from
0 to 9 twice. The samples were scanned to a 16x16 pixel grayscale
image. Each pixel was then thresholded to a Boolean value \cite{semeion}.
For experiments on this dataset, $\vec{y}^{(u)}$ and $\vec{y}^{(p)}$
are construct by grouping digits in the following ways:
\begin{itemize}
\item In the first experiment, the objective is to recognize digits 0 to
4 and protect digits 5 to 9. Utility, $\vec{y}^{(u)}$, is equal to
0 to 4 if the image is of such digit and it is equal to 5 if the image
is of 5, 6, 7, 8 or 9. Similarly, privacy label, $\vec{y}^{(p)}$,
is equal to 5 to 9 if the image is of such digit and it is equal to
0 otherwise. 
\item For the second experiment, the values of the $\vec{y}^{(u)}$ and
$\vec{y}^{(p)}$ are swapped. 
\end{itemize}
\begin{table}
\begin{centering}
\begin{tabular}{>{\centering}p{1.7cm}>{\centering}p{1.1cm}>{\centering}p{1.8cm}>{\centering}p{1.8cm}}
\toprule 
Label & Random Guess  & Before\\ Desensitization & After\\ Desensitization\tabularnewline
\midrule
\midrule 
{\small{}Activity (Utility)} & 16.67\% & 97.62\% & 92.48\%\tabularnewline
\midrule 
{\small{}Person Identification (Privacy)} & 5.26\% & 69.67\% & 7.02\%\tabularnewline
\bottomrule
\end{tabular}
\par\end{centering}
\caption{Results from the HAR dataset. \label{tab:results_har}}
\end{table}

\section{Results}

For all results, three accuracies are reported for comparison. The
random guess is the accuracy when no training is performed and the
prediction, hence, is made based on the frequency of the class in
the dataset. The accuracy before desensitization is resulted from
the prediction using full dimension of RDCA for the corresponding
label, along with the classifier. Finally, the accuracy after desensitization
is resulted from using the classifier on the desensitized data.

Table \ref{tab:results_har} reports the classification results on
the HAR dataset. Table \ref{tab:results_cmu} reports the results
from both experiments on the CMU Faces dataset. Finally, Table \ref{tab:results_hand_1}
and Table \ref{tab:results_hand_2} report the results from the two
experiments on the Semeion Handwritten Digit dataset. 

\begin{table}
\begin{centering}
\textbf{\small{}Experiment I}
\par\end{centering}{\small \par}
\begin{centering}
\begin{tabular}{>{\centering}p{1.2cm}>{\centering}p{1.1cm}>{\centering}p{1.8cm}>{\centering}p{1.8cm}}
\toprule 
Label & Random Guess  & Before\\ Desensitization & After\\ Desensitization\tabularnewline
\midrule
\midrule 
Pose (Utility) & 25.00\% & 83.30\% & 83.25\%\tabularnewline
\midrule 
Glasses (Privacy) & 50.00\% & 86.00\% & 50.47\%\tabularnewline
\bottomrule
\end{tabular}
\par\end{centering}
\begin{onehalfspace}
\begin{centering}
\textbf{\small{}Experiment II}
\par\end{centering}{\small \par}
\begin{centering}
\begin{tabular}{>{\centering}p{1.2cm}>{\centering}p{1.1cm}>{\centering}p{1.8cm}>{\centering}p{1.8cm}}
\toprule 
Label & Random Guess  & Before\\ Desensitization & After\\ Desensitization\tabularnewline
\midrule
\midrule 
Glasses

(Utility) & 50.00\% & 86.00\% & 85.97\%\tabularnewline
\midrule 
Pose (Privacy) & 25.00\% & 83.30\% & 25.00\% \tabularnewline
\bottomrule
\end{tabular}
\par\end{centering}
\end{onehalfspace}
\caption{Results from the two experiments on the CMU Faces dataset. \label{tab:results_cmu}}
\end{table}

\begin{table}
\begin{onehalfspace}
\begin{centering}
\textbf{Experiment I}
\par\end{centering}
\begin{centering}
Utility: 0-4
\par\end{centering}
\begin{centering}
\begin{tabular}{>{\centering}p{1.2cm}>{\centering}p{1.1cm}>{\centering}p{1.8cm}>{\centering}p{1.8cm}}
\toprule 
Digit & Random Guess  & Before\\ Desensitization & After\\ Desensitization\tabularnewline
\midrule
\midrule 
0 & 10.0\% & 95.61\% & 92.86\%\tabularnewline
\midrule 
1 & 10.0\% & 84.10\% & 73.71\%\tabularnewline
\midrule 
2 & 10.0\% & 86.30\% & 80.80\%\tabularnewline
\midrule 
3 & 10.0\% & 76.70\% & 72.50\%\tabularnewline
\midrule 
4 & 10.0\% & 83.14\% & 75.33\%\tabularnewline
\midrule 
The Rest & 50.0\% & 94.67\% & 92.46\%\tabularnewline
\bottomrule
\end{tabular}
\par\end{centering}
\begin{centering}
Privacy: 5-9
\par\end{centering}
\begin{centering}
\begin{tabular}{>{\centering}p{1.2cm}>{\centering}p{1.1cm}>{\centering}p{1.8cm}>{\centering}p{1.8cm}}
\toprule 
Digit & Random Guess  & Before\\ Desensitization & After\\ Desensitization\tabularnewline
\midrule
\midrule 
5 & 10.0\% & 90.16\% & 0.00\%\tabularnewline
\midrule 
6 & 10.0\% & 82.40\% & 0.00\%\tabularnewline
\midrule 
7 & 10.0\% & 85.24\% & 0.00\%\tabularnewline
\midrule 
8 & 10.0\% & 83.50\% & 0.00\%\tabularnewline
\midrule 
9 & 10.0\% & 88.30\% & 0.00\%\tabularnewline
\midrule 
The Rest & 50.0\% & 68.90\% & 99.86\%\tabularnewline
\bottomrule
\end{tabular}
\par\end{centering}
\end{onehalfspace}
\caption{Results from the first experiment on the Semeion Handwritten Digit
dataset, when the digits 0-4 are defined as the utility, whereas the
digits 5-9 are defined as the privacy. \label{tab:results_hand_1}}
\end{table}

\section{Discussion}

\subsection{The Effects of Desensitization on Privacy and Utility}

Five experiments on the three datasets indicate that desensitization
by RDCA can effectively protect privacy with respect to the privacy
label. The privacy accuracies drop to the random guess level in both
HAR and CMU Faces datasets, while the privacy accuracies of the privacy-sensitive
digits are almost zero in the Semeion Handwritten Digit dataset. Note
that the reason the privacy accuracies approach zero for the privacy-sensitive
digits is because the classifier predicts most samples to be in the
``don't care'', ``The Rest'', class, which is desirable for privacy
under the scenario considered.

On the other hand, desensitization does not attenuate the utility
as significantly. It reduces the utility accuracies of the HAR experiment
and both experiments on CMU Faces by only 5.14\%, 0.05\%, and 0.03\%,
respectively. On the Semeion Handwritten Digit experiments, the utility
accuracies also only drop by 7.53\% on average across all utility-relevant
digits. This shows that desensitization can be a viable tool in effectively
protecting privacy, while still providing good utility.

\begin{table}
\begin{onehalfspace}
\begin{centering}
\textbf{Experiment II}
\par\end{centering}
\begin{centering}
Utility: 5-9
\par\end{centering}
\begin{centering}
\begin{tabular}{>{\centering}p{1.2cm}>{\centering}p{1.1cm}>{\centering}p{1.8cm}>{\centering}p{1.8cm}}
\toprule 
Digit & Random Guess  & Before\\ Desensitization & After\\ Desensitization\tabularnewline
\midrule
\midrule 
5 & 10.0\% & 90.16\% & 75.20\%\tabularnewline
\midrule 
6 & 10.0\% & 82.40\% & 80.20\%\tabularnewline
\midrule 
7 & 10.0\% & 85.24\% & 81.70\%\tabularnewline
\midrule 
8 & 10.0\% & 83.50\% & 82.90\%\tabularnewline
\midrule 
9 & 10.0\% & 88.30\% & 64.90\%\tabularnewline
\midrule 
The Rest & 50.0\% & 68.90\% & 86.50\%\tabularnewline
\bottomrule
\end{tabular}
\par\end{centering}
\begin{centering}
Privacy: 0-4
\par\end{centering}
\begin{centering}
\begin{tabular}{>{\centering}p{1.2cm}>{\centering}p{1.1cm}>{\centering}p{1.8cm}>{\centering}p{1.8cm}}
\toprule 
Digit & Random Guess  & Before\\ Desensitization & After\\ Desensitization\tabularnewline
\midrule
\midrule 
0 & 10.0\% & 95.61\% & 0.01\%\tabularnewline
\midrule 
1 & 10.0\% & 84.10\% & 0.01\%\tabularnewline
\midrule 
2 & 10.0\% & 86.30\% & 0.00\%\tabularnewline
\midrule 
3 & 10.0\% & 76.70\% & 0.02\%\tabularnewline
\midrule 
4 & 10.0\% & 83.14\% & 0.01\%\tabularnewline
\midrule 
The Rest & 50.0\% & 94.67\% & 100.00\%\tabularnewline
\bottomrule
\end{tabular}
\par\end{centering}
\end{onehalfspace}
\caption{Results from the second experiment on the Semeion Handwritten Digit
dataset, when the digits 5-9 are defined as the utility, whereas the
digits 0-4 are defined as the privacy. \label{tab:results_hand_2}}

\end{table}

\subsection{Unique-Unique Privacy Problem}

One other variant of the privacy problem is Unique-Unique Privacy
problem. However, because all unique labels are surrogates for identity,
and differ in name only, when the objective is to protect a unique
label while trying to predict another unique label, the problem is
a contradiction. 

\subsection{Future Works }

RDCA approach to privacy should be extended to include regression.
This extension would be useful in a case where the utility is predicting
how much someone would be willing to spend on a house while privacy
is his savings account balance. Another useful extension is making
this method applicable to cases with multiple utility and privacy
labels \cite{al2017ratio}. For example predicting favorite activity
and food while protecting citizenship status and political affiliation.
With those two extensions, it would be interesting to try making all
variables in the dataset private except for the utility label or labels.
Then it may be possible to have a machine learning service in which
the desensitized data the user is submitting cannot be used to learn
the original data.

\section{Conclusion}

We defined three privacy problem classes in machine learning for classification,
in which the common goal is to maximize the classification accuracy
of the utility label, $\vec{y}^{(u)}$, while minimizing the classification
accuracy of the privacy label, $\vec{y}^{(p)}$. Common-Unique Privacy
problems have one label which is unique to each user. Common-Common
Privacy problems have both labels which are not unique to users. Split-Label
Privacy problems have $\vec{y}^{(u)}$ and $\vec{y}^{(p)}$ derived
from a single label $\vec{y}$ where some classes are grouped.

Based on five experiments, we show that data desensitization by RDCA
can effectively protect privacy across all three problem classes.
On HAR and CMU Faces datasets, the use of desensitized data results
in random guess level accuracies for privacy label at a cost of 5.14\%
and 0.04\%, on average, drop in accuracy on the utility label. For
Semeion Handwritten Digit dataset, accuracies of the privacy-sensitive
digits are almost zero and the accuracies for utility-relevant digits
drop by 7.53\% on average. In all experiments, the tradeoffs between
privacy and utility maybe acceptable and warrant a further exploration
and development of this method. 

\section*{Acknowledgement}

This material is based upon work supported in part by the Brandeis
Program of the Defense Advanced Research Project Agency (DARPA) and
Space and Naval Warfare System Center Pacific (SSC Pacific) under
Contract No. 66001-15-C-4068. The authors wish to thank Professor
J. Morris Chang from Iowa State University for invaluable discussion
and assistances.

\bibliographystyle{IEEEbib}
\bibliography{sources}

\begin{thebibliography}{10}

\bibitem{dataBreach}
Privacy~Rights Clearinghouse,
\newblock ``Chronology of data breaches,''
  \url{http://www.privacyrights.org/data-breach}, 2016,
\newblock Accessed: 2016-08-17.

\bibitem{netflix_prize}
Arvind Narayanan and Vitaly Shmatikov,
\newblock ``How to break anonymity of the netflix prize dataset,''
\newblock {\em CoRR}, vol. abs/cs/0610105, 2006.

\bibitem{RefWorks:181}
Arvind Narayanan, Hristo Paskov, Neil~Zhenqiang Gong, John Bethencourt, Emil
  Stefanov, Eui Chul~Richard Shin, and Dawn Song,
\newblock ``On the feasibility of internet-scale author identification,''
\newblock in {\em 2012 IEEE Symposium on Security and Privacy}. 2012, pp.
  300--314, IEEE.

\bibitem{RefWorks:182}
Joseph~A. Calandrino, Ann Kilzer, Arvind Narayanan, Edward~W. Felten, and
  Vitaly Shmatikov,
\newblock ``" you might also like:" privacy risks of collaborative filtering,''
\newblock in {\em 2011 IEEE Symposium on Security and Privacy}. 2011, pp.
  231--246, IEEE.

\bibitem{RefWorks:183}
Michael Barbaro and Tom~Zeller Jr.,
\newblock ``A face is exposed for aol searcher no. 4417749,''
  \url{http://www.nytimes.com/2006/08/09/technology/09aol.html}, Aug 9, 2006
  2006.

\bibitem{RefWorks:89}
Rakesh Agrawal and Ramakrishnan Srikant,
\newblock ``Privacy-preserving data mining,''
\newblock in {\em ACM Sigmod Record}. 2000, vol.~29, pp. 439--450, ACM.

\bibitem{RefWorks:90}
Dakshi Agrawal and Charu~C. Aggarwal,
\newblock ``On the design and quantification of privacy preserving data mining
  algorithms,''
\newblock in {\em Proceedings of the twentieth ACM SIGMOD-SIGACT-SIGART
  symposium on Principles of database systems}. 2001, pp. 247--255, ACM.

\bibitem{RefWorks:93}
Keke Chen and Ling Liu,
\newblock ``Privacy preserving data classification with rotation
  perturbation,''
\newblock in {\em Data Mining, Fifth IEEE International Conference on}. 2005,
  p. 4 pp., IEEE.

\bibitem{RefWorks:94}
Kun Liu, Hillol Kargupta, and Jessica Ryan,
\newblock ``Random projection-based multiplicative data perturbation for
  privacy preserving distributed data mining,''
\newblock {\em Knowledge and Data Engineering, IEEE Transactions on}, vol. 18,
  no. 1, pp. 92--106, 2006.

\bibitem{RefWorks:155}
Nico Schlitter and Jorg Lassig,
\newblock ``Distributed privacy preserving classification based on local
  cluster identifiers,''
\newblock in {\em Trust, Security and Privacy in Computing and Communications
  (TrustCom), 2012 IEEE 11th International Conference on}. 2012, pp.
  1265--1272, IEEE.

\bibitem{RefWorks:98}
Hwanjo Yu, Xiaoqian Jiang, and Jaideep Vaidya,
\newblock ``Privacy-preserving svm using nonlinear kernels on horizontally
  partitioned data,''
\newblock in {\em Proceedings of the 2006 ACM symposium on Applied computing}.
  2006, pp. 603--610, ACM.

\bibitem{RefWorks:103}
Keng-Pei Lin and Ming-Syan Chen,
\newblock ``On the design and analysis of the privacy-preserving svm
  classifier,''
\newblock {\em Knowledge and Data Engineering, IEEE Transactions on}, vol. 23,
  no. 11, pp. 1704--1717, 2011.

\bibitem{RefWorks:154}
Valeria Nikolaenko, Udi Weinsberg, Sotiris Ioannidis, Marc Joye, Dan Boneh, and
  Nina Taft,
\newblock ``Privacy-preserving ridge regression on hundreds of millions of
  records,''
\newblock in {\em Security and Privacy (SP), 2013 IEEE Symposium on}. 2013, pp.
  334--348, IEEE.

\bibitem{chanyaswad2017compressive}
Thee Chanyaswad, J~Morris Chang, and Sun-Yuan Kung,
\newblock ``A compressive multi-kernel method for privacy-preserving machine
  learning,''
\newblock in {\em Neural Networks (IJCNN), 2017 International Joint Conference
  on}. IEEE, 2017, pp. 4079--4086.

\bibitem{RefWorks:184}
Sun-Yuan Kung,
\newblock ``Compressive privacy: From information$\backslash$/estimation theory
  to machine learning [lecture notes],''
\newblock {\em IEEE Signal Processing Magazine}, vol. 34, no. 1, pp. 94--112,
  2017.

\bibitem{ACM2016}
Sun-Yuan Kung, Thee Chanyaswad, J~Morris Chang, and Peiyuan Wu,
\newblock ``Collaborative pca/dca learning methods for compressive privacy,''
\newblock {\em ACM Transactions on Embedded Computing Systems (TECS)}, vol. 16,
  no. 3, pp. 76, 2017.

\bibitem{mlsp2016}
Thee Chanyaswad, J~Morris Chang, Prateek Mittal, and Sun-Yuan Kung,
\newblock ``Discriminant-component eigenfaces for privacy-preserving face
  recognition,''
\newblock in {\em Machine Learning for Signal Processing (MLSP), 2016 IEEE 26th
  International Workshop on}. IEEE, 2016, pp. 1--6.

\bibitem{kung2015discriminant}
Sun-Yuan Kung,
\newblock ``Discriminant component analysis for privacy protection and
  visualization of big data,''
\newblock {\em Multimedia Tools and Applications}, pp. 1--36, 2015.

\bibitem{RefWorks:33}
S.~Y. Kung,
\newblock {\em Kernel Methods and Machine Learning},
\newblock Cambridge University Press, Cambridge, UK, 2014.

\bibitem{RefWorks:169}
Davide Anguita, Alessandro Ghio, Luca Oneto, Xavier Parra, and Jorge~Luis
  Reyes-Ortiz,
\newblock ``A public domain dataset for human activity recognition using
  smartphones.,''
\newblock in {\em ESANN}, 2013.

\bibitem{cmuFaces}
Tom Mitchell,
\newblock ``Cmu face images data set,''
  \url{https://archive.ics.uci.edu/ml/datasets/CMU+Face+Images}, 1997,
\newblock Accessed: 2016-9-3.

\bibitem{semeion}
``Semeion handwritten digit data set,'' Semeion Research Center of Sciences of
  Communication, via Sersale 117, 00128 Rome, Italy Tattile Via Gaetano
  Donizetti, 1-3-5,25030 Mairano (Brescia), Italy.

\bibitem{al2017ratio}
Mert Al, Shibiao Wan, and Sun-Yuan Kung,
\newblock ``Ratio utility and cost analysis for privacy preserving subspace
  projection,''
\newblock {\em arXiv preprint arXiv:1702.07976}, 2017.

\end{thebibliography}

\end{document}